\documentclass[twocolumn,aps,pre,superscriptaddress]{revtex4-1}

\usepackage{amssymb,amsmath,bm}
\usepackage{graphicx}

\usepackage{hyperref}
\usepackage{braket}

\providecommand{\abs}[1]{\vert#1\vert}

\providecommand{\e}{\text{e}} 
\providecommand{\dx}{\text{d}x}

\def\normOrd#1{\mathop{:}\nolimits\!#1\!\mathop{:}\nolimits}
 
\usepackage{color,soul}

\begin{document}
\title{An interacting adiabatic quantum motor}

\author{Anton Bruch}
\affiliation{\mbox{Dahlem Center for Complex Quantum Systems and Fachbereich Physik, Freie Universit\"at Berlin, 14195 Berlin, Germany}}
\author{Silvia Viola Kusminskiy}
\affiliation{Max Planck Institute for the Science of Light, Staudtstr. 2, 91058 Erlangen, Germany}

\author{Gil Refael}
\affiliation{
\mbox{Institute for Quantum Information and Matter, Caltech, Pasadena, CA 91125, USA}}

\author{Felix von Oppen}
\affiliation{\mbox{Dahlem Center for Complex Quantum Systems and Fachbereich Physik, Freie Universit\"at Berlin, 14195 Berlin, Germany}}

\date{\today}
\begin{abstract}
We present a field theoretic treatment of an adiabatic quantum motor. We explicitly discuss a motor termed Thouless motor which is 
based on a Thouless pump operating in reverse. When a sliding periodic potential is considered as the motor degree of freedom, a bias voltage applied to the electron channel sets the motor in motion. We investigate a Thouless motor whose electron channel is modeled  as a Luttinger liquid. Interactions increase the gap opened by the periodic potential. For an infinite Luttinger liquid the coupling induced friction is enhanced by electron-electron interactions. When the LL is ultimately coupled to Fermi liquid reservoirs, the dissipation reduces to its value for a noninteracting electron system for a constant motor
  velocity. Our results can also be applied to a motor based on a nanomagnet coupled to a quantum spin Hall edge. 

\end{abstract}

\maketitle


\section{Introduction}

Modern life is influenced by an immense variety of motors in all forms and sizes, driven by energy sources ranging from heat, as in a combustion engine, through chemical energy in biological motors to electrical energy in electric motors. As the miniaturization of modern devices moves towards ever smaller scales, the need for control over mechanical motion at these scales becomes increasingly pressing. Directed nanomechanical motion was realized using chemical energy \cite{Collins2016,Wilson2016}, light \cite{koumura1999,Klok2008}, and electrons \cite{Tierney2011,Kudernac2011} as driving agents.

We focus here on nanodevices, in which the motion of slow mechanical degrees of freedom is controlled by their coupling to electronic transport through the device -- forming a nano-electromechanical motor. One model for such a motor is based on an electron pump operating in reverse \cite{Qi2009,Bustos-Marun2013,Fernandez-Alcazar2015,Fernandez-Alcazar2017}. In such a pump, the cyclic variations of parameters, here effected by a mechanical degree of freedom, lead to a net charge transport through the device  \cite{Brouwer1998}. In the reverse mode a \textit{dc} bias is applied and the forces exerted by the scattered electrons drive the coupled mechanical degree of freedom \cite{BodePRL, Bode2012, Thomas2012, Thomas2015}, realizing a motor. 

\begin{figure}[b]
 
\includegraphics[width=7.0 cm,keepaspectratio]{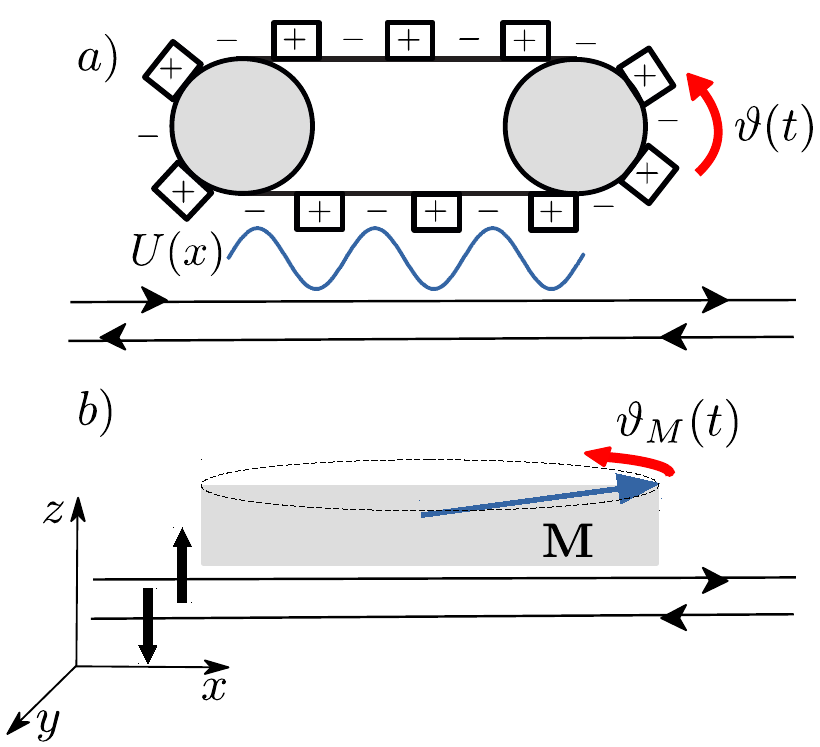}
\caption{\label{fig1} a) Model for a Thouless motor based on a single channel quantum wire in proximity to a chain of alternating charges. The sliding periodic potential $U(x)$ is associated with the rotational degree of freedom $\vartheta(t)$ of the quantum motor. b) A nanomagnet with magnetization $\bf M$ is coupled to a single edge of a quantum spin Hall insulator in the x-y-plane, where $\vartheta_M(t)$ (angle of the in-plane magnetization) is associated with the motor degree of freedom. }
\end{figure}

As an example, consider a device, in which electrons in a one-dimensional (1d) wire are coupled to a slowly sliding periodic potential, which is associated with the mechanical degree of freedom of the motor as depicted in Fig.\ \ref{fig1}. This device exhibits the essential features of the ancient Archimedean screw. When operating the Archimedean screw as a pump, turning of 
the screw leads to water transport. The same happens in the electronic system, where the sliding periodic potential pumps electrons through the 1d conductor, forming a Thouless pump \cite{thouless1983quantization}. When the Archimedean screw is operated in reverse, the water pushed through makes it work as a turbine. Similarly, in the electronic system a current pushed through the 1d conductor by an applied \textit{dc} bias voltage slides the periodic potential
associated with the slow mechanical degree of freedom, turning the device into a Thouless motor \cite{Qi2009,Bustos-Marun2013}.

Possible physical realizations of the Thouless motor were proposed based on a nanoscale helical wire placed in between capacitor plates \cite{Qi2009} and on  a quantum spin Hall (QSH) edge coupled to a nanomagnet \cite{Meng2014,Arrachea2015a,Silvestrov2016}. In the case of the helical wire, a slowly rotating transverse electric field leads to charge pumping, while in the inverse mode an applied \textit{dc} bias in presence of a static field leads to a rotation of the helix.
Similarly the precession of the magnetization of the nanomagnet pumps charge along the QSH edge, while in the inverse mode an applied bias leads to a spin transfer torque acting on the nanomagnet and driving its precession, cf. Fig.\ \ref{fig1}b).

The earlier theoretical description of adiabatic quantum motors assumed noninteracting electrons. When the electrons are confined to 1d, as in the present case of the quantum wire, the low energy behavior is modified by electron-electron interactions in essential ways. In this paper we investigate how these interaction effects modify the dynamics of adiabatic quantum motors. We describe the 1d electronic system as a Luttinger liquid (LL), which provides an exact description of its low energy excitations in terms of bosonic collective excitations \cite{giamarchi2004quantum}. 
LL theory has proven a useful description of both quantum wires \cite{Fisher1997,Auslaender2005} and QSH edges \cite{Wu2006}, covering the possible physical realizations of the Thouless motor mentioned above. Furthermore, our LL approach leads to a field theoretic description of quantum motors, complementing the earlier analysis on the basis of Landauer-B\"uttiker theory \cite{Bustos-Marun2013}. For definiteness we base our discussion on the Thouless motor and give an explicit translation of the results to the magnetic system in Sec.\ \ref{Translation}.

We introduce the model of the Thouless pump in Sec.\ \ref{Model}. In Sec.\ \ref{CouplingPeriodPot} we investigate the coupling of the LL to the periodic potential and derive the effective gap size in presence of electron-electron interactions. 
Section \ref{RedDyn} is devoted to the derivation of the effective field theory of the motor degree of freedom that 
leads to an interaction dependent effective Langevin equation for the motor dynamics. In the case of an infinite LL the friction is enhanced by repulsive interactions, as shown in Sec.\ \ref{infiniteLL}. The connection to Fermi liquid (FL) leads yields an effective equation of motion including memory and restores the reduced noninteracting dissipation at steady velocity, as presented in Sec.\ \ref{ContactFL}. In the final section \ref{Translation} we give the explicit translation of the obtained results to the nanomagnet coupled to a QSH edge.

\section{Model}\label{Model}
Our model of a quantum motor is based on
 a finite length Thouless pump operating in reverse. A toy model realizing such a pump is sketched in Fig.\ \ref{fig1}. A single
 channel quantum wire is placed next to a chain of fixed, periodically alternating
 charges. These charges move with respect to the quantum wire when turning the wheel and advancing 
 the angular degree of freedom $\vartheta$. This causes a slowly sliding periodic potential
 for the electrons, thereby forming
 a Thouless pump \cite{thouless1983quantization}.  

 The sliding periodic potential $U$ (cf. Fig.\ \ref{fig1}) is of the form 
\begin{equation}
U(x)=2\, V_0 \cos\left(q\, x-\vartheta(t)\right)\,\Theta\left(\frac{L}{2}-\abs{x}\right)\,,\label{eq:U}
\end{equation}
where $q$ is the wavevector of the potential of strength $2V_0$. For $q\approx 2k_{F}$ ($k_F$ is the Fermi momentum), the periodic potential causes backscattering between right and left moving electrons in the wire. 
The analysis of the system on the basis of Landauer-B\"uttiker theory for noninteracting electrons showed that, to exponential accuracy in the length $L$,  the backscattering induced gap leads to a vanishing normal conductance, quantized charge pumping per cycle, and unit efficiency, i.e., a conversion of the entire electronic energy provided by the bias into  mechanical energy associated with the degree of freedom $\vartheta$ \cite{Bustos-Marun2013}.

To include the interaction effects when confining the electrons to the 1d quantum wire, we model the electrons as a spinless  LL \cite{Haldane1981,giamarchi2004quantum}. The Hamiltonian
of the bare electronic system (i.e.\ without the periodic potential) can then be expressed in terms of the bosonic
displacement field $\phi(x)$ and phase field $\theta(x)$, 
\begin{equation}
H=\frac{v_{c}}{2\pi}\int d x\left\{ \frac{1}{K}\left(\partial_{x}\phi(x)\right)^{2}+K\left(\partial_{x}\theta(x)\right)^{2}\right\} \,,\label{eq:H Fields LL-1}
\end{equation}
where $K$
is the dimensionless interaction parameter, with $K<1$ for repulsive electron-electron interactions ($K=1 $ for a noninteracting system), and $v_{c}$
is the charge velocity.
The displacement field $\phi(x)$ describes
the local density fluctuations through 
\begin{equation}\label{NormOrdDensityPhi}
\normOrd{n_{R}(x)+n_{L}(x)}=\frac{\partial_{x}\phi(x)}{\pi} 
\end{equation}
and the phase field $\theta(x)$ is associated with the difference
in density between right and left movers,
\begin{equation}\label{LocalDiffernenceTheta}
\normOrd{n_{R}(x)-n_{L}(x)}=\frac{\partial_{x}\theta(x)}{\pi}\,.
\end{equation}
Here, $n_R$ and $n_L$ are the densities of right and left movers, respectively, and $\normOrd{\,...\,}$ denotes normal ordering.
The bosonic fields fulfill the commutation relation $[\phi(x),\theta(x')]=i\pi\,\text{sgn}(x-x')/2$. One can express the fermionic fields in terms of the bosonic ones via
\begin{align}
 \psi (x)&= \psi_R (x)+ \psi_L (x) \label{LeftRightMover}\,,\\
 \psi_{R/L}(x)&= \frac{1}{\sqrt{2\pi\lambda}} \text{e}^{\pm i  k_{F} x} \text{e}^{i \left[\theta(x) \pm \phi(x)\right]}\,, \label{PsiBosonized}
\end{align}
where we ignore the Klein factors and $\lambda$ is a short distance cutoff due to the finite band width \cite{giamarchi2004quantum}.

The Euclidean (imaginary time) action of the bare LL in the $\phi$-representation takes the form \cite{Fisher1997,kane1992transport}
\begin{equation}
S_0=\int d{\bf r}\frac{1}{2\pi K}\left[\frac{1}{v_{c}}\left(\partial_{\tau}\phi\right)^{2}+v_{c}\left(\partial_{x}\phi\right)^{2}\right]\label{eq:Phi-Repr}
\end{equation}
in terms of the short hand notations $(x,\tau)=\bf r $ and   $\int_{0}^{\beta} d \tau\int d x=\int d{\bf r}$. 
Using the bosonized fermionic fields in Eqs.\ \eqref{LeftRightMover} and \eqref{PsiBosonized}, the sliding periodic potential in Eq.\ \eqref{eq:U} contributes the sine-Gordon term
\begin{align}
S_{U}  =  \frac{2V_0}{2\pi\lambda}\int d{\bf r}\,\text{cos}\left[2\phi(x)+(2k_{F}-q)\, x+\vartheta(t)\right]\,\label{H_U}
\end{align}
for $x\in [-L/2,L/2]$ to the action.

\section{Coupling to periodic potential} \label{CouplingPeriodPot}

\subsection{Energy gap}\label{CalcEnergyGap}
The unit efficiency of the Thouless motor depends crucially on the presence of an energy gap at the Fermi energy. In the absence of interactions, this gap has size $\Delta_\text{non-int.} =2V_0$. Interactions modify this gap. To start with, the sine-Gordon term is a relevant perturbation over a wide range of interaction strengths, indicating the formation of a gap. 
Consider $\vartheta(t)=0$
and perfect backscattering, $q=2k_{F}$, and employ the usual momentum shell renormalization group (RG) procedure
for the sine-Gordon term in Eq.\ \eqref{H_U} \cite{giamarchi2004quantum}. Integrating out the fast modes of the action $S_0+S_U$ in Eqs.\ \eqref{eq:Phi-Repr} and \eqref{H_U} in a momentum shell  $\gamma/b<\abs{q}<\gamma$ ($\gamma$ is the momentum cutoff, see Appendix \ref{RGperfectBS}) and rescaling time $\tau'=\tau/b$ and space $x'=x/b$, with $b=\e^l$, yields the familiar flow equation 
\begin{equation}
\frac{ d V(l)}{ d l}=\left(2-K\right)V(l)\,,\label{eq:Delta-RG}
\end{equation}
for the strength of the periodic potential, while the free action $S_{0}$ remains unchanged to first order in the cumulant expansion. Thus the periodic potential is a relevant perturbation for all $K<2$ and the system flows to strong coupling.

\begin{figure} 	
 	\includegraphics[width=8.5cm,keepaspectratio]{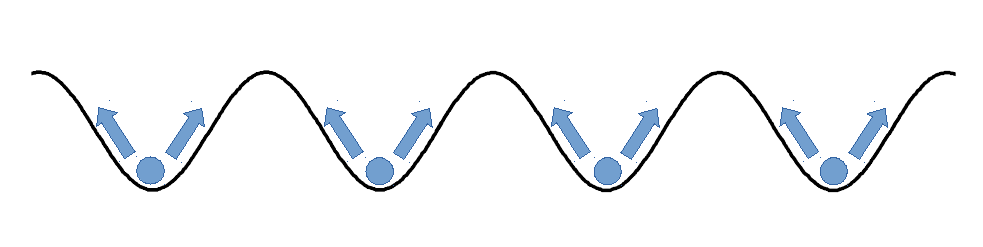}
 	\caption{\label{WignerCrystal} Oscillations of the electron density around the minima of the periodic potential corresponding to the action Eq.\ \eqref{SMassivePhase}. Quantum fluctuations of the electrons around the minima positions lead to a down-scaling of the strength of the  periodic potential $V$ as described by Eq.\ \eqref{eq:Delta-EffektiveLowEnergy}, resulting in the renormalized gap given by Eq.\ \eqref{eq:DeltaInteractingFinal}.}
 \end{figure}

For a large coupling strength $V$, the displacement field $\phi$ is trapped near a minimum of the cosine.
The electron density is commensurate and oscillates about the minima of the periodic potential in Fig.\ \ref{WignerCrystal}.
  The effective dynamics of $\phi$ can be obtained by expanding the action about this minimum,
\begin{align}
S  &\simeq  S_0 + \int d{\bf r}\frac{2V_0}{2\pi\lambda}\,\phi^{2} \nonumber \\
&= \sum_{n,m}\frac{1}{2\pi K}\left(\frac{1}{v_{c}}\omega_{n}^{2}+v_{c}q_{m}^{2}+\frac{4V_0K}{\lambda}\right)\abs{\phi_{n,m}}^{2} \,,\label{SMassivePhase}
\end{align}
where $\omega_n$ is a bosonic Matsubara frequency and $q_m$ is the wave vector. Thus, the system has a bare energy gap of size 
\begin{equation}\label{Delta bare}
\Delta_0=\sqrt{\frac{4V_0K\, v_{c}}{\lambda}}\,, 
\end{equation}
which can be understood as the pinning frequency of the classical Wigner crystal. Indeed, expanding the bare potential Eq.\ \eqref{eq:U} (with wavevector $2k_F$ and $\vartheta=0$) around a minimum yields $U\simeq  V_{0}\left[2k_{F}x\right]^{2}$. This leads to a pinning frequency
\begin{align}
\omega_\text{pin} = \sqrt{ \frac{8k_{F}^{2}}{m}V_{0}}\,.
\end{align}
Using $v_c K\simeq v_F $, this reproduces the bare gap $\Delta_0$ in Eq.\ \eqref{Delta bare} up to a numerical prefactor stemming from the uncertainty in choosing the cutoff $\lambda\sim 2 \pi/k_F $.

Quantum fluctuations of the electron density about the commensurate configuration (cf. Fig.\ \ref{WignerCrystal}) effectively decrease the restoring force of the potential and thus result in a downscaling of the effective gap. This effect is present even for noninteracting electrons. Indeed, for noninteracting electrons $K\rightarrow1$, the bare gap in Eq.\ \eqref{Delta bare} is different from $\Delta_\text{non-int.} =2V_0$. Repulsive interactions suppress the density fluctuations so that the downward renormalization becomes weaker as the repulsive interactions increase. In the Wigner crystal limit $K\rightarrow0$ fluctuations are fully suppressed, so that the bare gap Eq.\ \eqref{Delta bare} represents the actual gap of the system.

We account for the quantum fluctuations by integrating out the high energy modes while retaining the original units, so energies can be compared. This leads to 
\begin{equation}
\frac{ d V(l)}{ d l}=-KV(l)\,.\label{eq:Delta-EffektiveLowEnergy}
\end{equation}
We can see that, as anticipated, the downwards scaling is stronger for less repulsively interacting systems. 

Integrating out modes down to the gap leads to a self-consistent equation for the renormalized energy gap $\Delta=\sqrt{4V K\, v_{c}/\lambda}$, with $V$ obtained by integrating the flow equation \eqref{eq:Delta-EffektiveLowEnergy}, 
\begin{align}\label{VFlow}
V  =  V_{0}\left(\frac{2 \pi v_{c}}{ \lambda \Delta }\right)^{-K}\,.
\end{align}
The resulting self-consistent equation 
for $\Delta$ has the solution
\begin{equation}
\Delta =\left(\frac{4V_{0}K\, v_{c}}{\left(2 \pi v_{c} \lambda ^{-1} \right)^{K}\lambda}\right)^{1/(2-K)}\,.\label{eq:DeltaInteractingFinal}
\end{equation}
This formula reproduces $\Delta_\text{non-int.} = 2 V_0 $ for noninteracting
electrons (up to a numerical prefactor, as before). We also see explicitly that the gap is enhanced for repulsive electron-electron interactions
($K<1$),
\begin{align}\label{GapEnhancement}
\frac{\Delta(K)}{\Delta\left(K=1\right)} 
=  K \left(\frac{\pi^2 v_c/\lambda }{2V_{0}K}\right)^{(1-K)/(2-K)}>1\,.
\end{align}
Here we used that $\pi v_c/\lambda\gg V_0$ is an energy of the order of the Fermi energy. 

\subsection{Changes of the chemical potential}\label{deviationsPerfectBS}

The previous section considered the case of perfect commensurability 
$q=2k_{F}$ at the center of the gap $\mu=0$. The noninteracting Thouless motor maintains optimal
efficiency as long as the chemical potential falls into the gap
$\abs{\mu}\lesssim V_{0}$ \cite{Bustos-Marun2013}. We now  investigate the robustness of the interacting system against
changes of the chemical potential.

A uniform chemical potential term $H_{\mu}=-\mu\int dx\,\partial_{x}\phi(x)/\pi$
can be absorbed into the free LL Hamiltonian Eq.\ \eqref{eq:H Fields LL-1}
by shifting the field
\begin{align}
\tilde{\phi}(x)=\phi(x)-\mu\frac{K}{v_{c}}x\,.
\end{align}
This 
changes the coupling in Eq.\ \eqref{H_U} to \footnote{The electronic dispersion is linearized around $\pm k_F=\pm q/2$. Hence $2k_F-q=0$ in Eq.\ \eqref{H_U}.}
\begin{align}
S_{U}\left[\phi\right]=\frac{2V}{2\pi\lambda}\int d{\bf r}\,\text{cos}\left[2\tilde{\phi}(x)+2\mu\frac{K}{v_{c}}\,x\right]\,.\label{S-Deviation}
\end{align}
The chemical potential $\mu$ thus introduces a constant gradient $\nabla \tilde \phi =-\mu K/v_c$ 
into the configurations of $\tilde{\phi}$ that minimize the sine-Gordon term. Physically, this reflects the fact that the Luttinger liquid tries to adapt to a density which is commensurate with the periodic potential. The Luttinger liquid Hamiltonian in Eq.\ \eqref{eq:H Fields LL-1}
gives the associated elastic energy cost per unit length 
\begin{align}
\epsilon_{el}=\frac{v_{c}}{2\pi K}\left(\mu\frac{K}{v_{c}}\right)^{2}\,. \label{ElasticEnergy}
\end{align}
This cost increases with $\mu$ and eventually leads to depinning beyond a critical $\mu_{c}$, when adapting to the periodic potential becomes too costly. 

To take proper account of the renormalization of the potential due
to quantum fluctuations, we use the effective low energy theory developed in Sec.\ \ref{CalcEnergyGap}. Since
$\mu$ does not alter the renormalization of the strength of the periodic potential (up to first order in the cumulant expansion), we can express the
effective potential $V$ in terms of
the effective gap size $\Delta=\sqrt{4VK\,v_{c}/\lambda}$, with $\Delta$
given in Eq.\ \eqref{eq:DeltaInteractingFinal}. This leads to the
effective low energy action
\begin{align}
S_{\text{eff}}[\tilde{\phi}]=S_{0}[\tilde{\phi}]+\int d{\bf r}\frac{\Delta^{2}}{4\pi Kv_{c}}\,\text{cos}\left[2\tilde{\phi}(x)+2\mu\frac{K}{v_{c}}\,x\right]\,.\label{SeffDeviations}
\end{align}

The elastic energy cost in Eq.\ \eqref{ElasticEnergy} can be reduced
by inserting a finite density $n_{s}$ of $\pi$ phase slips into $\tilde{\phi}$, which are described by soliton solutions of $\tilde{\phi}$. With phase slips, the gradient of $\tilde \phi$ is no longer constant and has a reduced magnitude on average. We approximate the elastic energy cost $\epsilon_{el}$  of this configuration by calculating  $\epsilon_{el}$  associated with the reduced average gradient, which yields 
\begin{align}
\epsilon_{el}=\frac{v_{c}}{2\pi K}\left(\pi n_{s}-\mu\frac{K}{v_{c}}\right)^{2}\,.
\end{align}
With the assumption of a low soliton density the total energy cost can then be estimated as the sum of the elastic energy cost
and the cost of $n_{s}$ solitons, 
\begin{align}
\epsilon=\epsilon_{el}+n_{s}E_{sol}\,.\label{ETotSol}
\end{align}
The soliton solution and its energy $E_{sol}=2\Delta/(\pi K)$ can
be derived from Eq.\ \eqref{SeffDeviations} in the standard way
\cite{Rajaraman1987}. We
find the optimal soliton density by minimizing the total energy cost for a given chemical potential $\mu$,
\begin{align}
n_{s,\text{opt}}=\frac{\mu K}{\pi v_{c}}-\frac{2\Delta}{\pi^{2}v_{c}}\,.
\end{align}
This soliton density becomes positive at the critical chemical potential
\begin{align}
\mu_{c}=\frac{2}{\pi}\frac{\Delta}{K}\,,\label{DeltaCritical}
\end{align}
beyond which the system leaves the pinned regime. Since repulsive 
electron-electron interactions 
enhance the effective gap size $\Delta$ according to Eq.\ \eqref{eq:DeltaInteractingFinal},
 they also increase the robustness of the
system against changes of the chemical potential. Note that the
limit $K\rightarrow1$ of vanishing electron-electron interactions 
reproduces the critical chemical potential $\mu_{c}(K=1)\sim V_{0}$
of the noninteracting case.

\subsection{Sliding periodic potential}\label{CouplingTimeDepPot}
So far, we considered the motor degree of freedom to be at rest and chose $\vartheta=0$. 
In the absence of interactions, the adiabatic variation of $\vartheta$ pumps a unit charge between the leads per cycle. The same occurs in the interacting system.
Restoring the motor degree of freedom $\vartheta(\tau)$ in Eq.\ \eqref{H_U}, the coupling to the periodic potential is 
\begin{align}\label{S-timeDepDeviation}
S_{U}\left[\phi\right]=\frac{2V}{2\pi\lambda}\int d{\bf r}\,\text{cos}\left[2\phi(x)+\vartheta(\tau)\right]\,.
\end{align}
 This introduces an explicit time dependence into the solutions $\phi_{\text{min}}$
  that minimize the cosine
\begin{align}\label{PinnedPhiDevBS}
\phi_{\text{min}}(x,\tau)=-\frac{\vartheta(\tau)}{2}
\end{align}
(up to a constant that  picks the specific minimum of the cosine). A time-dependent displacement field $\phi$ implies current flow. 
  Using the continuity equation, we obtain the current density 
 \begin{align}
 j(x,t)=-\frac{e}{\pi}\partial_{t}\phi(x,t)=\frac{e}{2\pi}\partial_{t}\vartheta(t)\,,
\end{align}
which describes pumping of a quantized charge 
\begin{equation}\label{PumpedCharge}
 Q_{P}=\int_{0}^{T}dt\, j(t)=e
\end{equation}
when advancing the periodic potential by one period.

The interaction-enhanced gap implies a larger range of validity of this adiabatic treatment. Comparing the kinetic term in the Lagrangian to the energy gain due to the gap formation, we conclude that the adiabatic approximation remains valid as long as  $\abs{\dot{\vartheta}}\ll \Delta$, where $\Delta$ is the renormalized gap of the interacting system.

\section{Reduced dynamics of the motor degree of freedom}\label{RedDyn}

\subsection{Bias voltage} \label{BiasVoltage}
\begin{figure} 	
	\includegraphics[width=8.0cm,keepaspectratio] {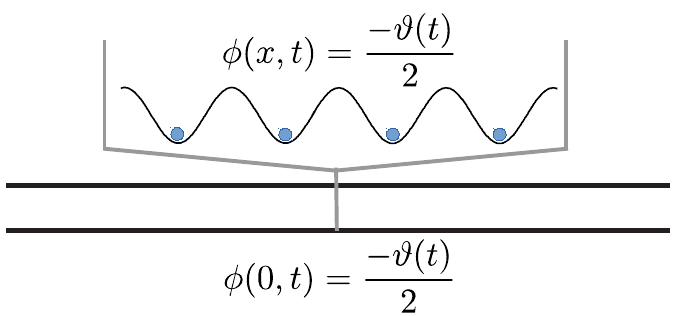}
	\caption{\label{SinglePoint} The pinning condition $\phi=-\vartheta/2$ within the area of the periodic potential reduces the coupling between motor degree of freedom and electrons to a free LL with a constrained boundary condition $\phi(0,t)=-\vartheta(t)/2$ when the area of the periodic potential is shrunk to a single point $x=0$.}
\end{figure}

As long as $\abs{\mu} < \mu_c$ and $\abs{\dot\vartheta} <\Delta$, the electrons within the region of the periodic potential are locked to the minima of the periodic potential (cf. Fig.\ \ref{SinglePoint}) and the electronic dynamics is effectively frozen out. This is reflected in a locked displacement field $\tilde \phi=-\mu K \, x/(2 v_c)-\vartheta/2$ and a gapped spectrum (from here on omit the tilde for notational simplicity). 
Effectively,  this allows us to shrink the length of the periodic potential to a single point $x=0$, at which the pinned displacement field $\phi(0,t)=-\vartheta(t)/2$ interacts with the free LLs, as shown schematically in Fig.\ \ref{SinglePoint}.

In a motor setup, the applied bias voltage $V$ is used to drive the motor degree of freedom $\vartheta (t)$. When the electronic dynamics in the region of the periodic potential is frozen out, the voltage can also be taken to drop at the point $x=0$. This yields a contribution to the action
\begin{align}\label{Bias}
	S_{\text{bias}}  =  -\frac{eV}{2}\int d{\bf r}\,\text{sgn}(x)\frac{\partial_x \phi}{\pi}=\frac{eV}{\pi}\int d\tau \phi(0,\tau )\,.
\end{align}
Here we used the bosonized form of the normal ordered electron density given in Eq.\ \eqref{NormOrdDensityPhi}. 

Integrating out all electronic degrees of freedom away from $x=0$ under the constraint $\phi(0,t)=-\vartheta(t)/2$, analogous to the treatment of a local impurity in a LL \cite{kane1992transport,CastroNetoAH1996}, leads to an effective description of the dynamics of the motor degree of freedom, including a non-conservative mean force stemming from the electronic bias, friction, and a fluctuating force.

\subsection{Motor dynamics for an infinite Luttinger liquid} \label{infiniteLL}
 We first treat the coupling to an infinite LL. Integrating out the LL (see \cite{kane1992transport} and App. \ref{SinglePointAction}), we obtain the effective action 
\begin{align}\label{SeffMatsubaraInfiniteLL}
 S_{\text{eff}}=\sum_{n}\left(\frac{{\cal I} \omega_{n}^2}{2}+ \frac{\abs{\omega_{n}}}{4\pi K} \right)\abs{\vartheta_{n}}^{2}-\int_{0}^{\beta} d \tau\frac{eV}{2\pi}\vartheta\,,
\end{align}
for $\vartheta (t)$. Here, we added the kinetic energy of the motor with its 
moment of inertia ${\cal I}$.
The second term describes a dissipative contribution to the motor dynamics and the third term a potential induced by the applied bias. 

To obtain the explicit equation of motion in real time, we analytically continue the action to the Keldysh contour \cite{KamenevFieldTheory}. The effective action then acquires the form
\begin{align}
S_{\text{eff}}&=\int\frac{ d \omega}{2\pi}(\bar{\vartheta}_{\omega}^{cl},\bar{\vartheta}_{\omega}^{q})\,\hat{K}(\omega)\,\left(\begin{array}{c}
\vartheta_{\omega}^{cl}\\
\vartheta_{\omega}^{q}
\end{array}\right)\,\,+\frac{eV}{\pi}\int d t\, \vartheta^q(t) \nonumber \\ 
\hat{K}(\omega)&=
\left(\begin{array}{cc}
0 & K^{A}(\omega)\\
K^{R}(\omega) & K^{K}(\omega)
\end{array}\right)\,.\label{KeldyshKernel}
\end{align}
We performed a Keldysh rotation of $\vartheta(t)$ into the quantum and classical components $\vartheta^{q}=(\vartheta^{+}-\vartheta^{-})/2$ and $\vartheta^{cl}=(\vartheta^{+}+\vartheta^{-})/2$, respectively. 
The kernels  $K^{R(A)}(\omega)$ are the analytical
continuations of the Matsubara correlator $\mathcal{K}(\omega_{n})={\cal I} \omega_{n}^2 /2+\abs{\omega_{n}}/(4\pi K) $ in Eq.\ \eqref{SeffMatsubaraInfiniteLL} to real
frequencies 
$K^{R(A)}(\omega)=-2\mathcal{K}(i\omega_{n}\rightarrow \omega \pm i \eta )$ \cite{KamenevFieldTheory}. The Keldysh component follows from the
fluctuation dissipation theorem, $K^{K}(\omega)=\left(K^{R}(\omega)-K^{A}(\omega)\right)\coth(\omega/2T)$ \footnote{Due to the pinned LL on the length of the periodic potential, the free LLs on the left and right side are isolated and act as independent equilibrium baths for the motor degree of freedom.}.
Fourier transforming the action Eq.\ \eqref{KeldyshKernel} to real time we
obtain 
\begin{align}
S=\int d t \,\bigg\{ & -2\vartheta^{q}(t)\left[{\cal I} \ddot{\vartheta}^{cl}(t)+\frac{\dot{\vartheta}^{cl}(t)}{2\pi K} -\frac{eV}{2\pi}\right]\nonumber \\
 & +\int dt'K^{K}(t-t')\vartheta^{q}(t)\vartheta^{q}(t')\bigg\} \,,\label{SeffRealTimeInfiniteLL}
\end{align}
where we performed an integration by parts. The Fourier transform
of the Keldysh component reads
\begin{equation}
K^{K}(t)=\frac{iT^{2}}{K\cosh^2\left(\pi Tt\right)}\,,
\end{equation}
yielding a coupling of the quantum fields which is nonlocal in time. This action determines the reduced dynamics of the motor degree of freedom including all quantum
fluctuations.

The contribution
quadratic in the quantum components leads to the fluctuating Langevin
force in the classical equation of motion of the motor. Its explicit form can be obtained by decoupling the quantum
components by a Hubbard-Stratonovich transformation \cite{KamenevFieldTheory}
\begin{align}
&\exp \left( i\int\frac{ d \omega}{2\pi}\, K^{K}(\omega)\abs{\vartheta^{q}(\omega)}^{2} \right)= \nonumber \\ 
&\int\mathbf{D}\left[\xi\right]\exp\left(\int\frac{ d \omega}{2\pi}\left[\frac{\abs{\xi(\omega)}^{2}}{i\, K^{K}(\omega)}+2i\bar{\xi}(\omega)\vartheta^{q}(\omega)\right]\right)\,, \label{HubbardStratLL}
\end{align}
where $\xi(t)$ is a real field. Introducing the integral over $\xi$ into the Keldysh partition function  $Z=\int\mathbf{D}\left[\vartheta\right]\exp(iS)$ with the action $S$ given in Eq.\ \eqref{SeffRealTimeInfiniteLL} 
leads to the classical saddle point equation for $\vartheta$
\begin{equation}
{\cal I} \ddot{\vartheta}^{cl}(t)=\frac{eV}{2\pi}-\frac{1}{2\pi K}\dot{\vartheta}^{cl}(t)+\xi(t)\label{EOMInfiniteLL}
\end{equation}
with 
\begin{equation}
\braket{\xi(t)\xi(t')}=\frac{K^{K}(t'-t)}{2i}=\frac{T^2}{2K\cosh^2\left(\pi T\left[t'-t\right]\right)}\,.
\end{equation}
Note that the friction coefficient $\gamma$ takes the value $\gamma=\hbar / 2\pi K$ when we reinsert $\hbar$. 

In the classical limit of large $T$ we can approximate $\cosh^{-2}\left(\pi T\left[t'-t\right]\right)\simeq 2(\pi T)^{-1}\delta(t-t')$
and the fluctuating force becomes $\delta$-correlated, with the magnitude of the correlator determined by temperature and the friction coefficient,
\begin{equation}
\braket{\xi(t)\xi(t')}=2\gamma T\delta(t-t')\,.
\end{equation}

We see that the mean force is unaffected by the electron-electron interactions, while the friction  $\gamma=(2 \pi K)^{-1}$ and with it the correlator of the fluctuating force are enhanced  by repulsive electron-electron interactions. For $K\rightarrow1$ the effective dynamics in Eq.\ \eqref{EOMInfiniteLL} reproduces the noninteracting result of Ref. \onlinecite{Bustos-Marun2013}. 

The time-averaged steady state velocity of the motor follows from the equation of motion \eqref{EOMInfiniteLL} which yields $\dot{\vartheta}=K\,eV$. 
Since the pumped charge in Eq.\ \eqref{PumpedCharge} is the only charge transported across the periodic potential, we can directly calculate the current $I$ as pumped charge per unit time 
\begin{align}\label{CurrentInfiniteLL}
 I=\frac{e\dot{\vartheta}}{2\pi}=\frac{K e^2}{ 2\pi \hbar}V\,,
\end{align}
where we reinserted $\hbar$ to bring the current into the usual form in terms of the conductance quantum. We can use this current at steady state to define the \textit{dc} conductance of the motor $g_M=K e^2/h$, which takes the value of an infinite, ideal LL \cite{kane1992transport}.

We use these results to investigate the efficiency $\eta$ of the interacting Thouless motor to perform work against an external load $F_{\,\text{load}}$. In the simple case that the load is independent of $\vartheta$, the steady velocity can be derived from Eq.\  \eqref{EOMInfiniteLL} via
\begin{align}\label{steadyVelocityWithLoad}
 \frac{ \dot \vartheta}{2\pi K}=\frac{eV}{2\pi}-F_{\,\text{load}}\,.
\end{align}
At this velocity, the work performed on the load per unit time takes the form
\begin{align}
 P_{\,\text{load}}=\dot \vartheta F_{\,\text{load}}=2\pi K \left(\frac{eV}{2\pi }-F_{\,\text{load}} \right) F_{\,\text{load}}\,.
\end{align}
This output power reaches a maximum
\begin{align}\label{MaximumPower}
 P_{\,\text{load,max}}=2\pi K \left(\frac{eV}{4\pi }\right)^2
\end{align}
at the load $F_{\,\text{load,max}}=eV/(4\pi)$.

The efficiency at maximum power is defined as the ratio between $P_{\,\text{load,max}}$ and the electrical input power $P_{in}=IV$ provided by the bias, which is determined by the pumped current in Eq.\ \eqref{CurrentInfiniteLL}. 
With the velocity at $F_{\,\text{load,max}}$, this yields an efficiency
\begin{equation}\label{EfficiencyInfiniteLL}
\eta=\frac{P_{\,\text{load,max}}}{P_{in}}=\frac{1}{4 \pi }\,
\end{equation}
at maximum power. We see that the maximum output power is reduced by repulsive electron-electron interactions ($K<1$) due to the increased dissipation. The reduced mean velocity for interacting systems also decreases the input power, which yields an interaction-independent efficiency at maximum power.

\subsection{Friction and energy current in an infinite Luttinger liquid}
It is interesting to obtain a more explicit description of the friction coefficient $\gamma$. To this end, we compute the energy current carried by the LL for the time dependent boundary condition $\phi(0,t)=-\vartheta(t)/2$. The solution of $\phi$ under this time dependent constraint is shown in  Eq.\ \eqref{RealTimeSolutionInfiniteLL} in Appendix \ref{SinglePointAction}. Assuming a steady velocity it takes the form 
\begin{align}\label{PhiSteadyVelocity}
\phi(x,t) =\frac{-\dot{\vartheta}}{2}\left(t-\frac{\abs{x}}{v_{c}}\right)\,.
\end{align}
To see how this solution carries the dissipated energy away from the motor, we investigate the energy current density $j^{E}$ corresponding to this solution. 
$j^{E}$ can be derived from the Heisenberg equation of motion for the energy density
\begin{align}
 \rho^{E}=\frac{v_{c}}{2\pi}\left[ \frac{1}{K}\left(\partial_{x}\phi(x)\right)^{2}+K\left(\partial_{x}\theta(x)\right)^{2}\right]\,,
\end{align} 
which yields
\begin{align}
\partial_{t}\rho^{E}  =  i\left[H,\rho^{E}\right] = -\frac{v_{c}^{2}}{2\pi}\partial_{x}\left\{ \partial_{x}\theta(x),\partial_{x}\phi(x)\right\} \,,
\end{align}
where we used the commutation relations of the bosonic fields introduced above in Sec.\ \ref{Model} and $ \{.,.\} $ denotes the anticommutator.
We can now directly deduce the energy current
via the continuity equation
\begin{equation}
\partial_{t}\rho^{E}  = -\nabla j^{E}
\end{equation}
which leads to
\begin{equation}\label{jE}
j^{E}=\frac{v_{c}^{2}}{2\pi}\left\{ \partial_{x}\theta(x),\partial_{x}\phi(x)\right\} \,.
\end{equation}
Since the gradient
of $\theta$ if fully determined by the time dependence of $\phi$, i.e., $\partial_{t}\phi(x,t)=i\left[H,\phi(x,t)\right]=-v_{c}K\partial_{x}\theta(x)$, we can write down the energy current corresponding to the solution in Eq.\ \eqref{PhiSteadyVelocity} as
\begin{equation}
j^{E}=
\frac{\dot{\vartheta}^{2}}{4\pi K} \text{sgn}(x)\,.
\end{equation}
Thus we can see that the dissipated power
\begin{equation}
-P_{diss}=\gamma\dot{\vartheta}^{2}=j^{E}(x>0)-j^{E}(x<0)\,
\end{equation}
is evenly split between the two sides and sent to $x=\pm \infty$.

\subsection{Contact to Fermi liquid leads} \label{ContactFL}
\begin{figure} 	
	\includegraphics[width=8.5cm,keepaspectratio] {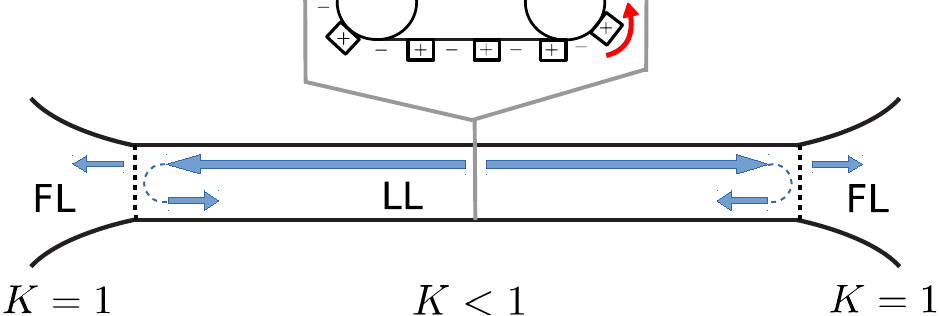}
	\caption{\label{LLFL}Connecting FL leads causes backscattering of plasmons at the FL-LL boundary.}
\end{figure}
In the previous section we assumed an infinite LL which leads to enhanced dissipation and a reduced motor conductance due to repulsive electron-electron interactions. It is well known that when contacting a LL by FL leads, the \textit{dc} conductance of the wire takes the value of an ideal noninteracting channel $g=e^2/h$ \cite{Maslov1995, Ponomarenko1995, Safi1995}. In this section we investigate, whether attaching FL leads reduces the dissipation of the Thouless motor to the noninteracting value and reproduces the noninteracting motor conductance $g_M=e^2/h$.

The FL leads generate backscattering of plasmons at the FL-LL boundary. This introduces memory into the effective equation of motion of the motor and results in reduced noninteracting dissipation at steady velocity, cf. Fig.\ \ref{LLFL}.
The transition between LL and FL can be modeled as a change of the interaction parameter $K\rightarrow1$ and an associated change of the charge velocity $v_c\rightarrow v_F$ \cite{Maslov1995,Karzig2011}.The LL is connected to FL reservoirs at $x=\pm D/2$, which yields 
\begin{align}
 S_0&= \int d{\bf r}\frac{1}{2\pi}\left[\frac{\left(\partial_{\tau}\phi\right)^{2}}{ K(x)\,v_c(x)}+\frac{v_c(x)\left(\partial_{x}\phi\right)^{2}}{K(x)}\right]\,,
\end{align}
for the action in the $\phi$-representation, with 
\begin{align}\label{SpaceDependentLL}
 K(x)=\begin{cases}
1 & \abs{x}\geq D/2\\
K & \abs{x}<D/2
\end{cases}
\end{align}
and
\begin{align}
v_{c}(x)=\begin{cases}
v_{F} & \abs{x}\geq D/2\\
v_{c} & \abs{x}<D/2\,.
\end{cases}
\end{align}
We again obtain the effective action of the motor by integrating out the electronic degrees of freedom under the constraint $\phi(0,t)=-\vartheta(t)/2$. The procedure amounts to solving the saddle point equation for the $\phi$-field in the presence of the appropriate boundary conditions for $\phi$ and $\partial_x\phi$, as shown in Appendix \ref{SinglePointAction}. This yields the effective action 
\begin{align}\label{SeffMatsubaraLLFL}
 S[\vartheta]_{\text{eff}}=\sum_{n}\frac{M(\omega_n)}{4\pi K}\abs{\omega_{n}}\,\abs{\vartheta_{n}}^{2}-\int_{0}^{\beta} d \tau\frac{eV}{2\pi}\vartheta\,,
\end{align}
with
\begin{align}\label{GammaLLFL}
M(\omega_n)=\left(1+2\sum_{n=1}^{\infty}\e^{-n\abs{\omega_n}\mathcal{T}}r_p^{n}\right)\,,
\end{align}
where $\mathcal{T} =D/ v_c $ is the traversal time of the plasmons from $x=0$ to the FL-LL boundary and back and $r_p=\frac{K-1}{K+1}$ is the plasmon reflection amplitude. 

To obtain the real time dynamics we analytically continue to the Keldysh contour analogous to the infinite LL case above. The kernels now take the form
\begin{align}
K^{R(A)}(\omega)&=\frac{\pm i\omega}{2\pi K}\left(1+2\sum_{n=1}^{\infty}\e^{\pm ni\omega \cal{T}}r_p^{n}\right)\,,        
\end{align}
and $K^K(\omega)=\left(K^{R}(\omega)-K^{A}(\omega)\right)\coth(\omega/2T)$ \cite{Note1}. Fourier transforming to real time shows that the plasmon scattering at the LL-FL boundary induces a coupling of the quantum field to earlier classical velocities,
\begin{align}
 S_{\text{diss}}= S_{qq}- &\int dt\, \frac{2\vartheta^{q}(t)}{2\pi K} \\
 \times& \left(\dot{\vartheta}^{cl}(t)+2\sum_{n=1}^{\infty}\dot{\vartheta}^{cl}(t-n\mathcal{T})r_p^{n}\right)\,.\nonumber 
\end{align}
Here
\begin{align}
 S_{qq}=\int d t  dt'K^{K}(t-t')\vartheta^{q}(t)\vartheta^{q}(t')
\end{align}
is the contribution of the dissipative action which is quadratic in the quantum fields.
In contrast to the infinite LL case above, the Keldysh kernel 
\begin{align}
K^K (t)= \frac{iT^{2}}{K}\left[\sum_{n=-\infty}^{\infty}\frac{1}{\cosh^{2}\left(\pi T\left[t+n{\cal T}\right]\right)}r_{p}^{\abs{n}}\right] 
\end{align}
leads to a nonlocal coupling of the quantum fields also in the high temperature limit, resulting in correlations of the fluctuating force which are nonlocal in time. As before we combine the dissipative action with the free part and the bias induced mean force and decouple the quantum fields via a Hubbard-Stratonovich transformation. This yields the nonlocal classical saddle point equation 
\begin{align}\label{EOMLLFL}
{\cal I} \ddot{\vartheta}^{cl}(t)&=\frac{eV}{2\pi}+\xi(t) \\
-&\frac{1}{2\pi K}\left[\dot{\vartheta}^{cl}(t)+2\sum_{n=1}^{\infty}\dot{\vartheta}^{cl}(t-n\mathcal{T})r_p^{n}\right]\,.\nonumber
\end{align} 
The correlator of the fluctuating force is given by
\begin{align}
 \braket{\xi(t)\xi(t')}=\frac{T^{2}}{2K}\sum_{n=-\infty}^{\infty}\frac{1}{\cosh^{2}\left(\pi T\left[t-t'+n{\cal T}\right]\right)}\,r_{p}^{\abs{n}}\,.
\end{align}
In the high temperature limit this leads to finite correlations at all multiples of the traversal time $\cal{T}$
\begin{align}
 \braket{\xi(t)\xi(t')}\simeq\frac{2T}{2\pi K}\sum_{n=-\infty}^{\infty}\delta(t-t'+n{\cal T})\, r_{p}^{\abs{n}}\,.
\end{align}

Since $r_p<0$, the nonlocal couplings to the velocity, i.e.\ the contribution $\propto r_p^n$ in Eq.\ \eqref{EOMLLFL} caused by multiple plasmon reflections at the FL-LL boundary and $x=0$, have alternating signs and a decaying amplitude $\propto \abs{r_p}^n$. Hence this force damps the motion for all even multiples of the traversal time and boosts the motion for all odd ones for a fixed sign of the velocity. How much energy is dissipated in this process depends on the detailed trajectory of $\vartheta$. 

At constant velocity, the effective dynamics in Eq.\ \eqref{EOMLLFL} leads to reduced dissipation and an enhanced velocity  $\dot{\vartheta}=eV$. This results in a larger pumped charge per unit time
\begin{align}
I=\frac{e\dot{\vartheta}}{2\pi}=\frac{e^2}{ 2\pi \hbar}V\,,
\end{align}
and hence  a \textit{dc} motor conductance which equals that of an ideal \textit{noninteracting} channel. Note that we again reinstated $\hbar$. Therefore, analogous to the \textit{dc} conductance of an ideal LL channel in contact to FL reservoirs \cite{Maslov1995}, also the \emph{dc motor conductance} is ultimately governed by the interactions in the attached reservoirs. Correspondingly the maximum output power of the Thouless motor is increased to the noninteracting value $K\rightarrow1$ in Eq.\ \eqref{MaximumPower}.

\subsection{Friction and energy current with attached Fermi liquid leads}
We now explore explicitly how the energy current is modified by plasmon reflections at the LL-FL boundary.
We consider two different trajectories: a constant velocity $\vartheta(t)=\dot{\vartheta}t$, and a sudden step $\vartheta(t)=\vartheta_{0}\Theta(t)$. For both cases, we derive the energy current as given above in Eq.\ \eqref{jE}, based on the solution for $\phi$ in Eq.\ \eqref{PhiRealtimeLLFL}.

For a sudden step, the gradient and time derivative of $\phi$  are strongly peaked $\delta$-functions that cannot interfere with each other. In this case, all the energy of the initial excitation is released into the FL reservoirs after integrating over multiple scattering events. Thus, the total dissipated energy 
\begin{align}
E_{diss}=\int dt [ j^{E}(x>0)-j^{E}(x<0) ] 
\end{align}
is determined by the initial plasmon excitation and takes the value of an infinite LL 
\begin{align}
E_{diss}=\frac{\tau \overline{\dot{\vartheta}^{2}}}{2\pi K}\,.
\end{align}
Here $\int dt \dot{\vartheta}(t)^{2}=\tau \overline{\dot{\vartheta}^{2}}$ determines the dissipation caused by the initial plasmon excitation and $\tau$ is the step duration.

In contrast for a constant velocity the reflected plasmons in Eq.\ \eqref{PhiRealtimeLLFL} interfere with each other, leading to a constant gradient $\partial_{x}\phi=K\dot{\vartheta}\,\text{sgn}(x)/(2v_{c})$ and time derivative $\partial_{t}\phi=-\dot{\vartheta}$ in the region $\abs{x}<D/2$.  This yields a reduced energy current  
\begin{align}\label{EnergyCurrentFLLeads}
j^{E}=\frac{\dot{\vartheta}^{2}}{4\pi}\text{sgn}(x)\,,
\end{align}
which corresponds to the dissipated power with the reduced \textit{noninteracting} friction $\gamma=(2\pi)^{-1}$
\begin{align}
-P_{diss}=\frac{1}{2\pi}\dot{\vartheta}^{2}=j^{E}(x>0)-j^{E}(x<0)\,.
\end{align}
Therefore it is interference of reflected plasmons (cf. Fig.\ \ref{LLFL}) which reduces the energy current at a constant velocity and thus prevents the system from releasing all the energy into the attached Fermi liquid leads.

\section{Magnetic motor}\label{Translation}

The counter-propagating states of a single QSH edge (cf. Fig \ref{fig1}) can be described as a Luttinger liquid analogous to the spinless quantum wire introduced in Sec.\ \ref{Model} \cite{Wu2006}. 
The bosonization of the helical channels is obtained from Eqs.\ \eqref{LeftRightMover} and \eqref{PsiBosonized} by replacing $\psi_{R}\rightarrow \psi_{R,\uparrow}$ and $\psi_{L}\rightarrow \psi_{L,\downarrow}$, while the Hamiltonian remains unchanged when written in terms of the bosonic fields. The exchange coupling to the nanomagnet $H_{M}=-J_0 /2 \int \dx \Psi^\dagger \pmb{\sigma} \Psi \cdot{\bf M}$ causes backscattering of the helical channels whenever $\bf M$ has a component in the x-y-plane. Here $\pmb{\sigma}$ is the vector of Pauli matrices and  $\Psi=(\psi_{R,\uparrow},\psi_{L,\downarrow})^T$. 
For strong easy-plane anisotropy, we can parametrize the magnetization as  $M_{x}=M \cos\vartheta_{M}$ and $M_{y}=M \sin\vartheta_{M}$, and the exchange coupling generates a sine-Gordon term
\begin{align}
 S_{M}=-\frac{ J_0 M} {2\pi\lambda}\int d {\bf r} \,\cos\left(2\phi(x)+2k_{F}x+\vartheta_{M}\right)\,.
\end{align}
Here $k_F$ is measured from the Dirac point $k=0$. Thus, the coupling of the nanomagnet to the helical edge states takes the same mathematical form as the coupling of the sliding periodic potential to the spinless LL in a quantum wire and we can directly translate the results of Sec.\ \ref{CouplingPeriodPot} and \ref{RedDyn} to the magnetic motor. For $K<2$ the $\phi$-field is locked to $\phi(x)=n\pi-k_{F}x-\vartheta_{M}/2$, which corresponds to alignment of the spin density along the exchange field of the nanomagnet
\begin{align}
 s_{x}(x)=\frac{1}{2\pi\lambda}\cos\vartheta_{M}\quad s_{y}(x)=\frac{1}{2\pi\lambda}\sin\vartheta_{M}\,.
\end{align}
A full precession of the magnetization leads to quantized charge pumping of one electron across the gapped region coupled to the nanomagnet. Quantum fluctuations lead to an interaction dependent downward scaling of the effective strength of the exchange coupling $J$, which results in an effective gap size for the lowest available modes 
\begin{equation}
\Delta_M =\left(\frac{2J_{0}M K\, v_{c}}{\left(2 \pi v_{c} \lambda ^{-1} \right)^{K}\lambda}\right)^{1/(2-K)}\,.\label{eq:DeltaMagnetic}
\end{equation}
This formula reproduces the gap $\Delta_\text{non-int.} = J_0 M $ for noninteracting helical edge modes and shows the strong enhancement of the magnetically induced gap by repulsive electron-electron interactions, cf. Eq.\ \eqref{GapEnhancement}. The noninteracting QSH edge remains insulating as long as the chemical potential remains within the gap that is opened by the magnet around the Dirac point $k=0$. Section \ref{deviationsPerfectBS} shows that interactions also make the magnetic system more robust against changes of the chemical potential and demonstrates that it remains gapped as long as $\abs{\mu}$ is smaller than $\mu_{c}=2 \Delta_M  /(\pi K) $, cf. Eq.\ \eqref{DeltaCritical}.

In the case of a large easy-plane anisotropy energy $D M^2_z/2$ ($D>0$), the Landau-Lifshitz-Gilbert equation governing the time evolution of the magnetization can be reduced to an equation of motion for the angle of the in-plane magnetization $\vartheta_M$, in which the inverse anisotropy constant acts as an effective moment of inertia ${\cal I}=D^{-1}$ \cite{Bode2012spin,Meng2014,Arrachea2015a}. With this we can readily translate the results for the effective dynamics Sec.\ \ref{RedDyn} to the magnetic case, replacing $\vartheta \rightarrow \vartheta_M$ and ${\cal I}\rightarrow D^{-1}$ . Thus, in the case of an infinite helical liquid, one obtains
\begin{align}
\frac{\ddot{\vartheta}_M^{cl}(t)}{D}=\frac{eV}{2\pi}-\frac{1}{2\pi K}\dot{\vartheta}_M^{cl}(t)+\xi(t)\,.
\end{align}
The dissipation is enhanced by repulsive interactions, leading to a reduced current and reduced motor conductance $g_M=K e^2/h$ compared to the noninteracting case in Eq.\ \eqref{CurrentInfiniteLL}. When assuming contact of the helical edge to Fermi liquid reservoirs as done in Sec.\ \ref{ContactFL}, the plasmon backscattering at 
the transition between helical liquid and reservoirs leads to an effective equation of motion including memory in Eq.\ \eqref{EOMLLFL} and the reduced dissipation of a noninteracting helical liquid at steady state.

\section{Summary}
\label{sec:Concl}
We investigated the effects of electron-electron interactions on a quantum motor that is based on a Thouless pump operating in reverse. Our field theoretic treatment enables a fully quantum description of the coupling induced motor dynamics. Repulsive interactions, of particular importance due to the reduced dimensionality of the system, enhance the energy gap opened by the coupling to the periodic potential. Interactions also increase the robustness of the system against changes in the chemical potential and increase the velocity range in which the system evolves adiabatically. Thus electron-electron interactions support the working principle of the motor. 

For infinite LLs with repulsive interactions, the friction experienced by the motor degree of freedom due to the coupling to the electrons is enhanced. When connecting the LL to FL reservoirs, plasmon reflections lead to an effective equation of motion including memory and a reduced dissipation which coincides with the noninteracting result for a constant 
motor velocity.  Consequently, the effective motor conductance is determined by the attached noninteracting reservoirs analogous to the \textit{dc} conductance of an ideal LL. 

Our result also applies to a nanomagnet coupled to the helical edge of a QSH system. This system can be readily mapped to the Thouless motor, possibly leading to an  experimentally more feasible realization of the motor.

\section*{Acknowledgments}

This work was supported in part by CRC 183 and SPP 1666 of the Deutsche Forschungsgemeinschaft. GR is grateful for support from the Packard Foundation, as well as from the IQIM, an NSF PFC, and to the Aspen Center for Physics, supported by the NSF grant PHY-1607761.
\bibliographystyle{apsrev4-1}
\bibliography{InteractAdiabatQMotor}


%

\appendix
\onecolumngrid

\section{RG perfect backscattering}\label{RGperfectBS}
We start with the LL action $S_0+S_U$  in the presence of the static  perfect backscattering potential, i.e. $\vartheta(t)=0$ and $q=2k_{F}$, in Eqs.\ \eqref{eq:Phi-Repr} and \eqref{H_U}. The calculation here is analogous to the RG treatment of the sine-Gordon term, e.g. in Ref. \onlinecite{giamarchi2004quantum}. We split the displacement field $\phi$ into slow and fast modes $\phi_{<}$ and $\phi{}_{>}$
\begin{equation}
\phi(\mathbf{r})=\phi_{<}(\mathbf{r})+\phi{}_{>}(\mathbf{r})\,,
\end{equation}
where $\phi_{<}$ contains frequencies and momenta inside the shell
$\abs{\abs{\mathbf{q}}}=\sqrt{k_{m}^{2}+(\omega_{n}/v_{c})^{2}}<\gamma/b$
with $b>1$ and $\gamma$ being the ultraviolet momentum cutoff. $\phi{}_{>}$ contains $\gamma/b<\abs{\abs{\mathbf{q}}}<\gamma$.
We integrate out the fast modes by averaging $\text{cos}\left[2\phi(x)\right]$
over the fast modes up to first order in the cumulant expansion
\begin{align}
\braket{S_{U}\left[\phi_{<},\phi{}_{>}\right]}_{0,>}=\int d{\bf r}\frac{V}{2\pi\lambda} \Big( \text{e}^{2i\phi_{<}(\mathbf{r})}\braket{\text{e}^{2i\phi_{>}(\mathbf{r})}}_{0,>} \nonumber +\text{h.c.} \Big)\,,
\end{align}
where $\braket{...}_{0,>}$ means averaging over the fast modes of
the free LL action, which is done in the Fourier decomposition
\begin{align}
\braket{\text{e}^{\pm2i\phi_{>}(\mathbf{r})}}_{0,>}=\frac{1}{Z_{0,>}}\prod_{n,m>}\int d \phi_{n,m}\exp\left\{ -\sum_{n,m>}\left(\frac{1}{2\pi K}\left(\frac{1}{v_{c}}\omega_{n}^{2}+v_{c}k_{m}^{2}\right)\abs{\phi_{n,m}}^{2}\pm\frac{2i}{\sqrt{\beta L}}\phi_{n,m}\text{e}^{i\mathbf{q}\cdot\mathbf{r}}\right)\right\} \,,
\end{align}
where $(n,m>)$ is a shorthand notation for the fast Fourier modes. Doing the Gaussian integral we obtain
\begin{align}
\braket{\text{e}^{\pm2i\phi_{>}(\mathbf{r})}}_{0,>}=\exp\left\{ -\sum_{n,m>}\frac{2K\pi}{v_{c}\,\beta\, L\,\left((\omega_{n}/v_{c})^{2}+k_{m}^{2}\right)}\right\} \,.
\end{align}
For low temperatures and large volumes the sum over the fast momentum
shell can be evaluated as a two-dimensional integral leading to
\begin{align}
\sum_{n,m>}\frac{2K\pi}{v_{c}\,\beta\, L\,\left((\omega_{n}/v_{c})^{2}+k_{m}^{2}\right)}=\frac{K}{2\pi}\int d \varphi\int d q\frac{q}{q^{2}}=K\,\ln(b)\,,
\end{align}
where $q=\abs{\mathbf{q}}$
and $\varphi=\text{arg}\left(k_{m}+i(\omega_{n}/v_{c})\right)$
is the angle in the $\mathbf{q}$
-plane. This leads to the effective action  
\begin{align}\label{SeffPerfectBS}
\braket{S_{U}\left[\phi_{<},\phi{}_{>}\right]}_{0,>}=\int d{\bf r}\frac{2V}{2\pi\lambda}\e^{-K\,\ln(b)}\cos\left[2\phi(\mathbf{r})\right]\,,
\end{align}
from which we derive the flow equation \eqref{eq:Delta-RG}.

\section{Effective action of the motor degree of freedom}

\label{SinglePointAction} In the following we derive the effective action
of the motor degree of freedom $\vartheta$ that results from the
coupling to the LL at the boundary of the periodic potential, given by the
pinning condition $\phi(0,t)=-\vartheta(t)/2$. Since the coupling
to an applied bias Eq.\ \eqref{Bias} $\propto\phi(0,\tau)$ is already
given in terms of the field at $x=0$ only, it is not affected by
the integrating out procedure. We can directly impose the constraint
$\phi(0,\tau)=-\vartheta(\tau)/2$ and obtain its contribution to
the effective action of $\vartheta$ 
\begin{align}
S_{\text{bias}}=-\int_{0}^{\beta}d\tau\frac{eV}{2\pi}\vartheta(\tau)\,.
\end{align}
To obtain the dissipative part of the effective dynamics, we integrate
out the fields of the LL under the constraint $\phi(0,\tau)=-\vartheta(\tau)/2$,
which we implement via a functional delta function 
\begin{equation}
\delta\left[\phi(0,\tau)+\frac{\vartheta(\tau)}{2}\right]=\int D[\kappa]\exp\left(-i\int_{0}^{\beta}d\tau\,\kappa(\tau)\left[\phi(0,\tau)+\frac{\vartheta(\tau)}{2}\right]\right)\,,\label{FunctionalDelta}
\end{equation}
where we introduced an additional real-valued field $\kappa(\tau)$.
With that we can write effective action $S_{\text{diss}}[\vartheta]$
as 
\begin{equation}
\exp(-S_{\text{diss}}[\vartheta])=\int D[\phi]\,\delta\left[\phi(0,\tau)+\frac{\vartheta(\tau)}{2}\right]\exp\left(-S_{0}\right)=\int D[\phi]\,D[\kappa]\,\exp\left(-S\right)\,.
\end{equation}
Using Eq.\ \eqref{FunctionalDelta}, the exponent $-S$ reads 
\begin{equation}
S=\frac{1}{2\pi}\int dr\left[\frac{1}{K(x)\,v_{c}(x)}\left(\partial_{\tau}\phi\right)^{2}+\frac{v_{c}(x)}{K(x)}\left(\partial_{x}\phi\right)^{2}\right]+i\int_{0}^{\beta}d\tau\,\kappa(\tau)\left(\phi(0,\tau)+\frac{\vartheta(\tau)}{2}\right)\,,\label{Sdelta}
\end{equation}
where the space-dependent charge velocity and interaction parameter
Eq.\ \eqref{SpaceDependentLL} model the contact to the FL leads.
We find the effective action by solving the saddle point equations
under a variation of $\phi$ and $\kappa$, which enables us to to
take proper account of the boundary conditions at the intersections
of FL and LL and the constraint $\phi(0,\tau)=-\vartheta(\tau)/2$
. For that we write down the saddle point equation under a variation
of $\phi$ 
\begin{eqnarray}
\frac{1}{\pi}\left(\frac{\omega_{n}^{2}}{K(x)\,v_{c}(x)}-\partial_{x}\frac{v_{c}(x)}{K(x)}\partial_{x}\right)\phi(x,\omega_{n}) & = & -i\kappa(\omega_{n})\delta(x)\,.\label{SaddlePointEq}
\end{eqnarray}
We can solve this equation by writing the solutions for each region
of constant $v_{c}$ and $K$ and solving for their coefficients by
taking proper account of the boundary conditions as shown below. Inserting
the implicit solution of the saddle point equation into the action
\eqref{Sdelta} leads to 
\begin{eqnarray}
S & = & \frac{i}{2}\int d\tau\kappa(\tau)\phi(0,\tau)+i\int_{0}^{\beta}d\tau\,\kappa(\tau)\frac{\vartheta(\tau)}{2}\,,\label{SdeltaImplicit}
\end{eqnarray}
where $\phi(0,\tau)$ is the solution of the saddle point equation.
From there we can later obtain the saddle point equation for $\kappa$
to reach the desired dissipative action of the motor.

\paragraph{Infinite Luttinger liquid}

We start with the infinite LL case. To find the explicit solution
for $\phi(0,\tau)$ we write down the solution of Eq.\ \eqref{SaddlePointEq}
for $x>0$ and $x<0$ 
\begin{equation}
\phi(x,\omega_{n})=\begin{cases}
\beta\e^{\frac{\abs{\omega_{n}}}{v_{c}}x} & x<0\\
\delta\e^{-\frac{\abs{\omega_{n}}}{v_{c}}x} & 0<x\,,
\end{cases}
\end{equation}
where we directly omitted the part of the solution that grows for
$x\rightarrow\pm\infty$. The appropriate boundary conditions follow
from the saddle point equation Eq.\ \eqref{SaddlePointEq} and are
given by the continuity $\phi(x=0^{+})=\phi(x=0^{-})$, which directly
gives $\beta=\delta$, and 
\begin{eqnarray}
\frac{v_{c}}{\pi K}\partial_{x}\phi(x,\omega_{n})\vert_{x=0^{-}}^{x=0^{+}} & = & i\kappa(\omega_{n})\label{BCx0}\\
\rightarrow\beta & = & -\frac{i\kappa(\omega_{n})\pi K}{2\abs{\omega_{n}}}=\phi(0,\omega_{n})\,.\label{betaOfKappa}
\end{eqnarray}
We use the solution for $\phi(0,\omega_{n})$ and insert it
into $S$ Eq.\ \eqref{SdeltaImplicit} 
\begin{equation}\label{SIntermediate}
S=\frac{1}{4}\sum_{\omega_{n}}\frac{\abs{\kappa(\omega_{n})}^{2}\pi K}{\abs{\omega_{n}}}+i\sum_{\omega_{n}}\left(\kappa(\omega_{n})\frac{\vartheta(-\omega_{n})}{2}\right)\,.
\end{equation}
A variation of $\kappa(\omega_{n})$ leads to the saddle point equation
\begin{eqnarray}
\kappa(-\omega_{n}) & = & -i\frac{\vartheta(-\omega_{n})\abs{\omega_{n}}}{\pi K}\,.\label{kappaOmega}
\end{eqnarray}
Inserting $\kappa$ Eq.\ \eqref{kappaOmega} into the action Eq.\ \eqref{SIntermediate} yields the dissipative action used in the main text Eq.\ \eqref{SeffMatsubaraInfiniteLL}
\begin{eqnarray}
S_{\text{diss}}[\vartheta] & = & \sum_{\omega_{n}}\frac{\abs{\omega_{n}}}{4\pi K}\abs{\vartheta_{n}}^{2}\,.\label{SDissInfiniteLL}
\end{eqnarray}
Furthermore we can use the solution for $\kappa(-\omega_{n})$ Eq.\ \eqref{kappaOmega}
to determine $\beta=-\vartheta(\omega_{n})/2$ in Eq.\ \eqref{betaOfKappa}
and write down the real space solution for $\phi$ 
\begin{equation}
\phi(x,\omega_{n})=\begin{cases}
-\frac{\vartheta(\omega_{n})}{2}\e^{\frac{\abs{\omega_{n}}}{v_{c}}x} & x<0\\
-\frac{\vartheta(\omega_{n})}{2}\e^{-\frac{\abs{\omega_{n}}}{v_{c}}x} & 0<x\,.
\end{cases}
\end{equation}
Analytical continuation to real frequencies $\abs{\omega_{n}}\rightarrow-i\omega$
and taking the Fourier transform to real time leads to 
\begin{eqnarray}
\phi(x,t) & = & \begin{cases}
-\frac{\vartheta\left(t+\frac{x}{v_{c}}\right)}{2} & x<0\\
-\frac{\vartheta\left(t-\frac{x}{v_{c}}\right)}{2} & 0<x\,.
\end{cases}\label{RealTimeSolutionInfiniteLL}
\end{eqnarray}
Thus, the boundary condition travels with $\pm v_{c}$
on the right (left) side.

\paragraph{Contact to Fermi liquid leads}

In the case of a finite LL in contact to FL leads, we need to write
down the solution of the saddle point equation \eqref{SaddlePointEq}
in the various regions 
\begin{equation}
\phi(x,\omega_{n})=\begin{cases}
A\e^{\frac{\abs{\omega_{n}}}{v_{F}}x} & x<-D/2\\
\left[\beta+\gamma/2\right]\e^{\frac{\abs{\omega_{n}}}{v_{c}}x}+\left[\beta-\gamma/2\right]\e^{-\frac{\abs{\omega_{n}}}{v_{c}}x} & -D/2<x<0\\
\left[\delta+\epsilon/2\right]\e^{\frac{\abs{\omega_{n}}}{v_{c}}x}+\left[\delta-\epsilon/2\right]\e^{-\frac{\abs{\omega_{n}}}{v_{c}}x} & 0<x<D/2\\
F\e^{-\frac{\abs{\omega_{n}}}{v_{F}}x} & D/2<x\,,
\end{cases}\label{PhiRegionsFLLL}
\end{equation}
where we again directly omitted the part of the solution that grows
for $x\rightarrow\pm\infty$. Additionally to the boundary condition
at $x=0$ Eq.\ \eqref{BCx0} we get at $x=D/2$ an additional condition
from the saddle point Eq.\ \eqref{SaddlePointEq} 
\begin{eqnarray}
-\frac{1}{\pi}\frac{v_{c}(x)}{K(x)}\partial_{x}\phi(x,\omega_{n})\vert_{x=0^{-}}^{x=0^{+}} & = & 0\\
\frac{v_{c}}{K}\partial_{x}\phi\left(\frac{D}{2}^{-},\omega_{n}\right) & = & v_{F}\partial_{x}\phi\left(\frac{D}{2}^{+},\omega_{n}\right)\,,
\end{eqnarray}
where $\phi$ has to be continuous and the analogous condition at
$x=-D/2$. Solving for the coefficient $\beta=\phi(0,\omega_{n})/2$
leads to 
\begin{align}
\phi(0,\omega_{n}) & =-\frac{i\kappa(\omega_{n})\pi K}{2\abs{\omega_{n}}\,M(\omega_{n})}\label{PhiKappaFLLL}\\
M(\omega_{n}) & =\frac{\left(1+\frac{1}{K}\right)\e^{\frac{\abs{\omega_{n}}D}{2v_{c}}}+\left(1-\frac{1}{K}\right)\e^{-\frac{\abs{\omega_{n}}D}{2v_{c}}}}{\left(1+\frac{1}{K}\right)\e^{\frac{\abs{\omega_{n}}D}{2v_{c}}}-\left(1-\frac{1}{K}\right)\e^{-\frac{\abs{\omega_{n}}D}{2v_{c}}}} \nonumber \\
&=\left(1+2\sum_{n=1}^{\infty}\e^{\frac{-n\abs{\omega_{n}}D}{v_{c}}}\left[\frac{K-1}{K+1}\right]^{n}\right)\nonumber \\
&=\left(1+2\sum_{n=1}^{\infty}\e^{-n\abs{\omega_{n}}\mathcal{T}}r_{p}^{n}\right)\label{M}\,,
\end{align}
where we used the plasmon or charge reflection amplitude $r_{p}=\frac{K-1}{K+1}$
and the traversal time of the plasmons from x=0 to the FL-LL boundary
and back $\mathcal{T}=D/v_{c}$. Using the explicit solution
for $\phi(0,\tau)$ in Eq.\ \eqref{SdeltaImplicit} yields
\begin{equation}
S=\frac{1}{4}\sum_{\omega_{n}}\frac{\pi K}{\abs{\omega_{n}}\,M(\omega_{n})}\abs{\kappa(\omega_{n})}^{2}+i\sum_{\omega_{n}}\left(\kappa(\omega_{n})\frac{\vartheta(-\omega_{n})}{2}\right)\,.\label{SimplicitKappa}
\end{equation}
A variation of $\kappa(\omega_{n})$ leads to 
\begin{eqnarray}
\kappa(-\omega_{n}) & = & -i\frac{\vartheta(-\omega_{n})\abs{\omega_{n}}M(\omega_{n})}{\pi K}\,,\label{kappaomegaFLLL}
\end{eqnarray}
which we insert into the action Eq.\ \eqref{SimplicitKappa} to obtain
the dissipative action used in the main text Eq.\ \eqref{SeffMatsubaraLLFL}
\begin{equation}
S_{\text{diss}}[\vartheta]=\sum_{\omega_{n}}\frac{\abs{\omega_{n}}\,M(\omega_{n})}{4\pi K}\abs{\vartheta_{n}}^{2}\,.
\end{equation}
Finally we use $\kappa(-\omega_{n})$ Eq.\ \eqref{kappaomegaFLLL}
to obtain the solution of $\phi$ under the constraint $\phi(0,t)=-\vartheta(t)/2$.
We focus here on the inner part $\abs{x}<D/2$, since the outer parts
do not provide any additional insight. Inserting $\kappa(-\omega_{n})$
Eq.\ \eqref{kappaomegaFLLL} into Eq.\ \eqref{PhiKappaFLLL} and
using the relations between the coefficients of $\phi$ Eq.\ \eqref{PhiRegionsFLLL}
that we obtained from matching the boundary conditions, we get 
\begin{align}
\phi(x,\omega_{n}) & =-\frac{\vartheta(\omega_{n})}{4}\begin{cases}
\left[1+M\right]\e^{\frac{\abs{\omega_{n}}}{v_{c}}x}+\left[1-M\right]\e^{-\frac{\abs{\omega_{n}}}{v_{c}}x} & -D/2<x<0\\
\left[1-M\right]\e^{\frac{\abs{\omega_{n}}}{v_{c}}x}+\left[1+M\right]\e^{-\frac{\abs{\omega_{n}}}{v_{c}}x} & 0<x<D/2
\end{cases}\\
\phi(x,\omega_{n}) & =-\frac{\vartheta(\omega_{n})}{4}\begin{cases}
\left[2+2\sum_{n=1}^{\infty}\e^{-n\abs{\omega_{n}}\mathcal{T}}r_{p}^{n}\right]\e^{\frac{\abs{\omega_{n}}}{v_{c}}x}-2\left[\sum_{n=1}^{\infty}\e^{-n\abs{\omega_{n}}\mathcal{T}}r_{p}^{n}\right]\e^{-\frac{\abs{\omega_{n}}}{v_{c}}x} & -D/2<x<0\\
-2\sum_{n=1}^{\infty}\e^{-n\abs{\omega_{n}}\mathcal{T}}r_{p}^{n}\e^{\frac{\abs{\omega_{n}}}{v_{c}}x}+\left(2+2\sum_{n=1}^{\infty}\e^{-n\abs{\omega_{n}}\mathcal{T}}r_{p}^{n}\right)\e^{-\frac{\abs{\omega_{n}}}{v_{c}}x} & 0<x<D/2\,,
\end{cases}
\end{align}
where we used Eq.\ \eqref{M}. Analytical continuation to real frequencies
$\abs{\omega_{n}}\rightarrow-i\omega$ and Fourier transformation
to real times leads to 
\begin{align}
\phi(x,t) & =-\frac{1}{2}\begin{cases}
\vartheta\left(t+\frac{x}{v_{c}}\right)-\sum_{n=1}^{\infty}\vartheta\left(t-\frac{x}{v_{c}}-{\cal T}n\right)r_{P}^{n}+\sum_{n=1}^{\infty}\vartheta\left(t+\frac{x}{v_{c}}-{\cal T}n\right)r_{P}^{n} & -D/2<x<0\\
\vartheta\left(t-\frac{x}{v_{c}}\right)-\sum_{n=1}^{\infty}\vartheta\left(t+\frac{x}{v_{c}}-{\cal T}n\right)r_{P}^{n}+\sum_{n=1}^{\infty}\vartheta\left(t-\frac{x}{v_{c}}-{\cal T}n\right)r_{P}^{n} & 0<x<D/2\,.
\end{cases}\label{PhiRealtimeLLFL}
\end{align}
Here we can see that additionally to the initial excitation that travels
with $\pm v_{c}$ to the right (left) side, $\phi$ contains the field
that gets reflected with amplitude $-r_{p}$ at the LL-FL boundary
and is subsequently reflected with amplitude $-1$ at $x=0$ and so
on, which leads to the reduced uniform energy current Eq.\ \eqref{EnergyCurrentFLLeads}
in the main text.

\end{document}